\documentclass[twocolumn, amsmath, amssymb, nofootinbib, superscriptaddress]{revtex4}

\usepackage{graphicx}
\usepackage{amsmath}
\usepackage{amssymb}
\usepackage{xcolor}

\pacs{78.67.Sc, 78.20.Bh, 73.20.Mf, 84.30.-r, 87.10.Hk}

\begin{document}
\title{Two-dimensional plasmons in the random impedance network model of disordered thin film nanocomposites}

\author{N.\,A.\,Olekhno}
\affiliation{ITMO University, 49 Kronverksky Pr. St. Petersburg, 197101 Russia}
\email{olekhnon@gmail.com}
\author{Y.\,M.\,Beltukov}
\affiliation{Ioffe Institute, Politekhnicheskaya ul. 26, St. Petersburg, 194021 Russia}

\date{\today}

\begin{abstract}
Random impedance networks are widely used as a model to describe plasmon resonances in disordered metal-dielectric nanocomposites. Two-dimensional networks are applied when considering thin films despite the fact that such networks correspond to the two-dimensional electrodynamics [J.P. Clerc {\it et al},  J. Phys. A {\bf 29}, 4781 (1996)]. In the present work, we propose a model of two-dimensional systems with the three-dimensional Coulomb interaction and show that this model is equivalent to the planar network with long-range capacitive links between distant sites. In the case of a metallic film, we obtain the well-known dispersion of two-dimensional plasmons $\omega \propto \sqrt{k}$. We study the evolution of resonances with a decrease in the metal filling factor within the framework of the proposed model. In the subcritical region with the metal filling $p$ lower than the percolation threshold $p_c$, we observe a gap with Lifshitz tails in the spectral density of states (DOS). In the supercritical region $p>p_c$, the DOS demonstrates a crossover between plane-wave two-dimensional plasmons and resonances of finite clusters.
\end{abstract}

\maketitle

\section{Introduction}

Disordered metal-dielectric films are nanocomposites whose structure depends on the surface filling fraction of a metal. Their geometry ranges from a dielectric surface with isolated metallic granules to a conducting metallic layer with dielectric holes~\cite{2000_Sarychev}. The local scale of such inclusions is tens of nanometers, however, they can form big metallic and dielectric clusters having a complex structure. Such systems have attracted great interest because of their optical properties, which are due to surface plasmons~\cite{Maier-book}. Plasmon resonances in disordered films form collective modes that are characterized by a complicated structure and giant fluctuations of local electric fields~\cite{2000_Sarychev}. These fluctuations are regions with extremely high local fields, hot spots, which play a crucial role in the process of surface-enhanced Raman scattering~\cite{1997_Brouers, Le_Ru_2009}. Other effects also increase in such disordered systems, especially near the percolation threshold, for example, high harmonic generation~\cite{2000_Sarychev, 2001_Breit}. Plasmon resonances modify the electromagnetic local density of states, thus allow controlling the rate of decay for excitations related to atoms, molecules, and other quantum emitters by exploiting the Purcell effect \cite{2015_Gaio, 2015_Carminati, 2016_Szilard}.

Various effective medium approaches have been developed~\cite{Choy_2016} in order to study macroscopic properties, such as optical absorption or electric conductivity of a film. However, these theories do not accurately describe the local geometry of a composite, and cannot therefore describe field fluctuations and the local density of states. To this end, other approaches were introduced, such as impedance network models~\cite{1990_Clerc}. The latter description follows from Maxwell's equations within the quasistatic approximation~\cite{1971_Kirkpatrick, 1973_Kirkpatrick, 1977_Webman}. This approximation can be used since the characteristic scale of inclusions is much smaller than the light wavelength.

A detailed study of two-dimensional systems has been performed within the framework of the random impedance network model. Optical absorption has been considered in Ref.~\cite{1987_Koss}. The spectral density of resonances has been studied for binary disordered systems~\cite{1998_Jonckheere} and fractal clusters~\cite{1996_Clerc}. Considerable attention was paid to localization properties of resonances~\cite{2006_Seal} and the statistics of local field fluctuations in disordered systems. It was shown that these fluctuations demonstrate a multifractal behavior~\cite{1998_Jonckheere}. In some cases, a mapping between resonances in random networks and the Anderson localization problem \cite{Abrahams} has been demonstrated~\cite{1999_Sarychev, 1999_Fyodorov}, which was used to study the localization in potentials with flat bands~\cite{2010_Chalker} as well as the localization of surface plasmons \cite{1999_Sarychev}. Various scaling laws have been established for physical quantities in systems near the percolation threshold~\cite{1991_Brouers, 1993_Brouers}. Resonances in ordered networks were also studied in order to explore properties of metasurfaces and photonic crystals~\cite{2004_Gu, 2005_Schafer}. The model was also applied to study various properties of three-dimensional composites: optical absorption~\cite{1989_Zeng, 1993_Zhang}, the spectral density of resonances~\cite{2000_Albinet}, and local field fluctuations~\cite{2004_Dai}.

In the mentioned papers, two-dimensional networks with a topology of the square lattice were used to describe properties of thin film composites. As we discuss further, the two-dimensional network model corresponds to the two-dimensional electrodynamics \cite{1996_Clerc}. Thus, this model should be modified in order to properly describe plasmon resonances in thin films.

The paper is organized as follows. In Sec.~\ref{sec:RIN}, we briefly recall the derivation of the impedance network model from Maxwell's equations. Kirchoff's rules for a network are reduced to a generalized eigenvalue problem, which we use to carry out numerical simulations in the present work. In Sec.~\ref{sec:reduced}, we analyze field distributions of elementary metallic clusters in a dielectric lattice and put forward a new impedance network model of thin film composites. We show that the composite should be replaced by the network with long-range capacitive links between distant sites rather than by the square lattice of impedances. In Sec.~\ref{sec:plasmons}, the model under consideration is applied to resonances in thin metal films. Resonances in the model of an infinitely thin metallic layer are shown to be two-dimensional plasmons. Effects related to the geometrical disorder are considered in Sec.~\ref{sec:crossover}. It is shown that Lifshitz tails in the DOS are present at metal fillings $p$ lower than the percolation threshold $p_c$, whereas at fillings $p>p_c$ the crossover is observed between two-dimensional plasmons and resonances of typical clusters. We also notice the presence of a smooth peak in the DOS of resonances in a disordered network divided by the DOS of the two-dimensional plasmon. Such a maximum is often found in amorphous solids and referred as the {\it boson peak}. Section~\ref{sec:conclusion} contains final remarks and the discussion of results.

\section{The random impedance network model}
\label{sec:RIN}

First, we briefly recall the derivation of the network model following the lines of works \cite{1971_Kirkpatrick, 1973_Kirkpatrick, 1977_Webman}. Since characteristic scales of metallic and dielectric inclusions are much smaller than the wavelength of visible and infrared radiation, the quasistatic approximation can be applied. In this case, the electric field is curl free ($\operatorname{rot}{\bf E}=0$), and the electrostatic potential $\varphi$ can be introduced (${\bf E} = -\operatorname{grad}\varphi$).

It is well known that Maxwell's equations can be reduced to the equation for an eddy current $\mathop{\rm div}\mathbf{j}=0$ within the quasistatic approach \cite{1973_Kirkpatrick, 1977_Webman}. At a given frequency $\omega$, the current $\mathbf{j}$ and the electric field $\mathbf{E}$ obey constitutive relation $\mathbf{j}(\omega, \mathbf{r})=\sigma(\omega, \mathbf{r})\mathbf{E}(\omega, \mathbf{r})$. The conductivity $\sigma(\omega, \mathbf{r})$ is related to the permittivity $\varepsilon(\omega, \mathbf{r})$ of the same region as $\sigma(\omega, \mathbf{r})=i\omega\varepsilon(\omega, \mathbf{r})/4\pi$ \cite{LL8}. In the considered binary nanocomposite, metallic regions with a permittivity $\varepsilon(\omega, \mathbf{r}) = \varepsilon_m(\omega)$ and dielectric regions with a permittivity $\varepsilon(\omega, \mathbf{r}) = \varepsilon_d(\omega)$ are present.

Next, we discretize the problem on a mesh with a topology of the simple cubic lattice with the lattice constant $a$. Equations ${\mathop{\rm div}{\bf j}=0}$ and ${\mathop{\rm rot}{\bf E}=0}$ are transformed to Kirchhoff's current rule and Kirchhoff's voltage rule, respectively. One can simultaneously represent both Kirchhoff's rules as a linear system
\begin{equation}
    \sum_j \left[\sigma_m(\omega)M_{ij} + \sigma_d(\omega)D_{ij}\right]\varphi_j = 0
    \label{eq:ls}
\end{equation}
with $\sigma_{m,d}(\omega, \mathbf{r})=i\omega\varepsilon_{m,d}(\omega, \mathbf{r})/4\pi$. Matrices $M$ and $D$ are Laplacian matrices (also known as Kirchhoff matrices) \cite{Bollobaas} of metallic and dielectric regions, respectively. Off-diagonal matrix elements are $M_{ij}=-1$ if sites $i$ and $j$ are connected by metallic impedance and $M_{ij}=0$ otherwise. Diagonal elements of the matrix $M$ are defined as $M_{ii}=-\sum_{j \neq i}M_{ij}$. The matrix $D$ has the same definition but for dielectric impedances. The sum of these matrices ${\cal L} = M+D$ is the Laplacian matrix of the regular simple cubic lattice. Thus, a disordered metal-dielectric composite can be treated as a binary random impedance network.

In the simplest case, the permittivity of metallic regions can be described within the Drude model $\varepsilon_m(\omega)=1-\omega_p^2/\omega^2$, and the permittivity of dielectric regions is constant: $\varepsilon_d(\omega) = \varepsilon_d$. In this case, ``metallic'' bonds are parallel $LC$ circuits with parameters
\begin{equation}
    \displaystyle L_m=\frac{4\pi c^2}{a\omega_p^2}, \quad C_m=\frac{a}{4\pi},
    \label{eq:metal}
\end{equation}
while ``dielectric'' bonds are capacitors with a capacitance
\begin{equation}
    \displaystyle C_d=\frac{\varepsilon_da}{4\pi}.
    \label{eq:dielectric}
\end{equation}
One can see that the resonant frequency of a single ``metallic'' bond is the plasma frequency of a metal $\omega_p = c/\sqrt{L_mC_m}$. The inductance $L_m$ is attributed to the kinetic energy of charge carriers in a metal instead of the energy stored in a magnetic field. Such an inductance plays a crucial role in metals at high frequencies $\omega\sim \omega_p$ and is known as the \emph{kinetic inductance}.

Resonant frequencies and corresponding eigenmodes in an arbitrary binary impedance network can be obtained as solutions to the generalized eigenvalue problem \cite{1998_Jonckheere}, which follows from the system of linear equations (\ref{eq:ls})
\begin{equation}
    \sum_jM_{ij}\varphi_j(\lambda_n) = \lambda_n \sum_j(M_{ij} + D_{ij})\varphi_j(\lambda_n).
    \label{eq:GEP}
\end{equation}
Eigenvalues $\lambda_n$ are related to resonant frequencies $\omega_n$ as
\begin{equation}
    \lambda_n = \frac{\sigma_d(\omega_n)}{\sigma_d(\omega_n) - \sigma_m(\omega_n)} = \frac{\varepsilon_d(\omega_n)}{\varepsilon_d(\omega_n) - \varepsilon_m(\omega_n)}.   \label{eq:lambda}
\end{equation}
Matrices $M$ and $D$ are positive semidefinite \cite{Horn-book}. Therefore, all of the eigenvalues satisfy the inequality $0 \le \lambda_n \le 1$ \cite{Ortega-book}. Eigenvectors $\varphi_j(\lambda_n)$ describe potentials at sites of the network. Eigenmodes of the problem (\ref{eq:GEP}) correspond to dielectric resonances in the model network \cite{1992_Bergman}, which represent plasmon resonances in the composite. The generalized eigenvalue problem (\ref{eq:GEP}) does not depend on the dielectric functions $\varepsilon_m(\omega)$ and $\varepsilon_d(\omega)$. Thus, only Eq.~(\ref{eq:lambda}) determines resonant frequencies $\omega_n$, which can be obtained from known eigenvalues $\lambda_n$ and dielectric functions $\varepsilon_m(\omega)$ and $\varepsilon_d(\omega)$. This allows mapping of resonances in systems with the same geometry but different dielectric functions of constituents \cite{2016_OBP}. If the dielectric function of  a metal is taken in the form $\varepsilon_m(\omega)=1-\omega_p^2/\omega^2$, then resonant frequencies are in the range $0 \le \omega_n \le \omega_p$. Thus, all of the results in the present paper can be easily generalized to arbitrary functions $\varepsilon_m(\omega)$ and $\varepsilon_d(\omega)$.

Some results on the spectral properties of long-range one-dimensional random networks have been obtained analytically within the framework of the random matrix theory \cite{2001_Fyodorov, 2003_Staring}, but a numerical study is still the main tool which is used to analyze properties of resonances in random networks. In the present work, we use the Kernel polynomial method \cite{2006_Weisse, 2016_Beltukov} to resolve the problem (\ref{eq:GEP}) numerically for networks with a sufficiently large number of bonds.

\section{The reduced network model of thin-film composites}
\label{sec:reduced}

\begin{figure}[t]
    \includegraphics[width=8.5cm]{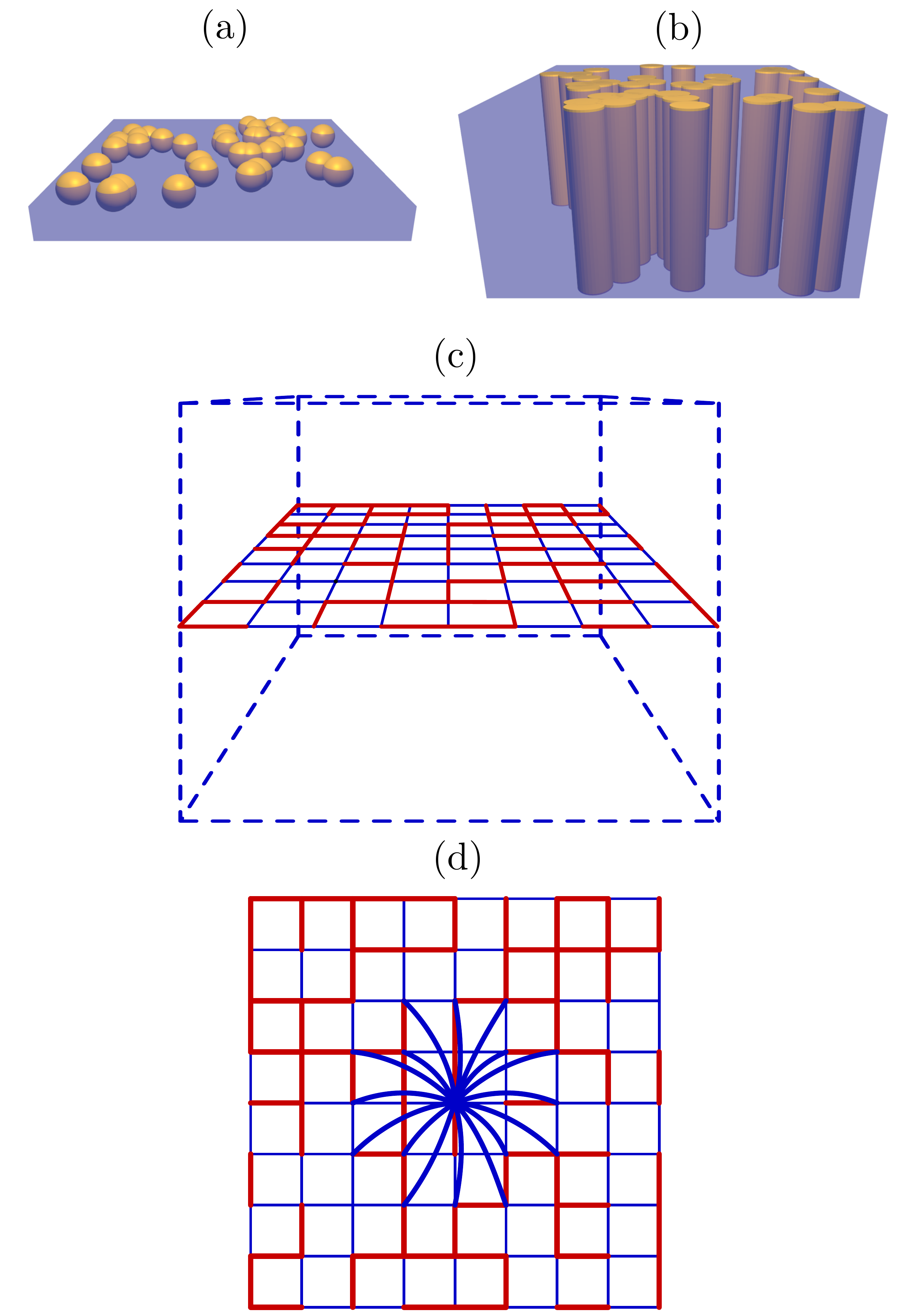}
    \caption{(Color online) (a) Sketch of a random thin film composite. (b) Example of a composite which corresponds to a two-dimensional network. (c) Three-dimensional random impedance network with $p=0.5$. Only the plane which contains metallic-type impedances as well as dielectric ones is shown. Thick red (black) lines denote metallic-type impedances, whereas thin blue (gray) lines denote dielectric-type impedances. (d) The equivalent two-dimensional network with long-range bonds (few of them are shown for the central site).}
    \label{fig:Film_and_nanowire}
\end{figure}

Two-dimensional impedance networks with a topology of the square lattice are commonly used for a theoretical study of plasmon resonances in thin film composites \cite{2000_Sarychev}. However, the vacuum in the considered quasistatic approach is equivalent to a dielectric medium with the permittivity $\varepsilon_d=1$ as it supports displacement currents. Thus, two-dimensional networks correspond to the two-dimensional electrodynamics \cite{1996_Clerc}, that is, to composites consisting of metallic nanowires which are placed into a dielectric medium (Fig.~\ref{fig:Film_and_nanowire}). Indeed, electrostatic Green's function for the square lattice impedance network has the following form \cite{1996_Clerc}:
\begin{equation}
    \displaystyle G(x) = \frac{1}{\pi}(2{\rm ln}(x) + {\rm ln}(8) + 2\gamma + O(1/|x|^2)),
\end{equation}
where $x=r_{ij}$ is the distance between a charge located at the site $i$ and an observation point at the site $j$, and $\gamma=0.57721$ is Euler's constant. This expression can be obtained within the random walk theory \cite{Spitzer-book}. The first term in the equation above corresponds to the potential of a point charge in the two-dimensional electrostatics, whereas omitted terms of relative order $1/|x|^2$ and higher describe anisotropic corrections which arise due to the lattice discreteness\cite{1996_Clerc}. On the other hand, Green's function for the three-dimensional simple cubic lattice demonstrates the asymptotic behavior $G(x)\propto 1/2\pi x$ \cite{Spitzer-book} that corresponds to the three-dimensional Coulomb interaction.

As a result, a more appropriate model for a thin film composite is an impedance network with a single layer (or, generally, a group of adjacent layers) containing ``metallic'' bonds as well as ``dielectric'' ones which is surrounded by the three-dimensional lattice of ``dielectric'' bonds (Fig.~\ref{fig:Film_and_nanowire}). Spectral properties of thin film composites, as well as the localization of resonances in such systems, should be reconsidered within this network model. The above statement doesn't affect results related to three-dimensional nanocomposites, e.g., those of papers \cite{1987_Koss, 1989_Zeng, 2000_Albinet}.

The three-dimensional lattice contains a big number of sites with $z\neq0$, which form the surrounding dielectric medium. They sufficiently increase a complexity of the eigenvalue problem (\ref{eq:GEP}) but do not increase the number of nontrivial resonances. However, the three-dimensional problem can be reduced to the two-dimensional one with a much smaller number of sites. Equation~(\ref{eq:GEP}) can be formally written as
\begin{equation}
    \sum_j{\cal L}_{ij}\varphi_j = I_i,   \label{eq:phi}
\end{equation}
where $I_i = \sum_jM_{ij}\varphi_j/\lambda_n$. Equation~(\ref{eq:phi}) can be treated as an equation for site potentials $\varphi_j$ in the conductive simple cubic lattice made of unit resistors and external currents $I_i$. $M_{ij}$ is nonzero only if $z_i=0$ and $z_j=0$ since all metallic bonds located in the layer $z=0$. Therefore, all nonvanishing external currents $I_i$ located in the plane $z_i=0$.

In order to solve the generalized eigenvalue problem~(\ref{eq:GEP}), only the knowledge of potentials in the plane $z_i=0$ is necessary. Therefore, one can find a linear relation between in-plane currents and in-plane potentials. This relation can be written as
\begin{equation}
    {\sum_j}'{\cal L}'_{ij}\varphi_j = I_i,   \quad z_i=0,   \quad z_j=0 \label{eq:phi2d}
\end{equation}
and should be valid for any $\varphi_j$ and $I_i$ in the plane $z=0$, which hold in the original problem (\ref{eq:phi}). The matrix elements ${\cal L}'_{ij}$ correspond to the bonds in the reduced two-dimensional network. The sum $\sum_j'$ denotes the summation over $j$ with the constraint $z_j=0$.

In the original three-dimensional problem~(\ref{eq:phi}), we have the simple cubic lattice with unit bonds between the nearest neighbors. Therefore, we can use the Fourier transform to solve Eq.~(\ref{eq:phi}) and find the in-plane relation~(\ref{eq:phi2d}). The resulting reduced network has a form of the square lattice with additional bonds
\begin{equation}
    {\cal L}'_{ij} = \int\sqrt{S^2(k_x, k_y) + S(k_x, k_y)}\,e^{i{\bf k}{\bf r}_{ij}}dk_xdk_y,
\end{equation}
where $S(k_x, k_y)=\sin^2(k_xa/2) + \sin^2(k_ya/2)$, ${\bf r}_{ij}$ is the distance between sites $i$ and $j$, and the integration is performed over the first Brillouin zone of the square lattice. In contrast to the original three-dimensional problem, the reduced two-dimensional network has nonzero elements between an arbitrary pair of sites at a distance ${\bf r}_{ij}$. One can show that ${\cal L}'_{ij}\sim r_{ij}^{-3}$.

Finally, the generalized eigenvalue problem~(\ref{eq:GEP}) is reduced to the equivalent two-dimensional problem
\begin{equation}
    {\sum_j}'M_{ij}\varphi_j = \lambda_n {\sum_j}'(M_{ij} + D'_{ij})\varphi_j, \quad z_i=0.
    \label{eq:GEP2d}
\end{equation}
The matrix elements $D'_{ij}={\cal L}'_{ij} - M_{ij}$ describe effective dielectric connections between in-plane sites $i$ and $j$. Therefore, long-range dielectric-type impedances are present in the reduced network [Fig.~\ref{fig:Film_and_nanowire}(d)] with \begin{equation}
    D'_{ij} \propto r_{ij}^{-3}.
\end{equation}

\section{Plasmonic eigenmodes in the reduced model}
\label{sec:plasmons}

\begin{figure}[t]
    \includegraphics[width=8.5cm]{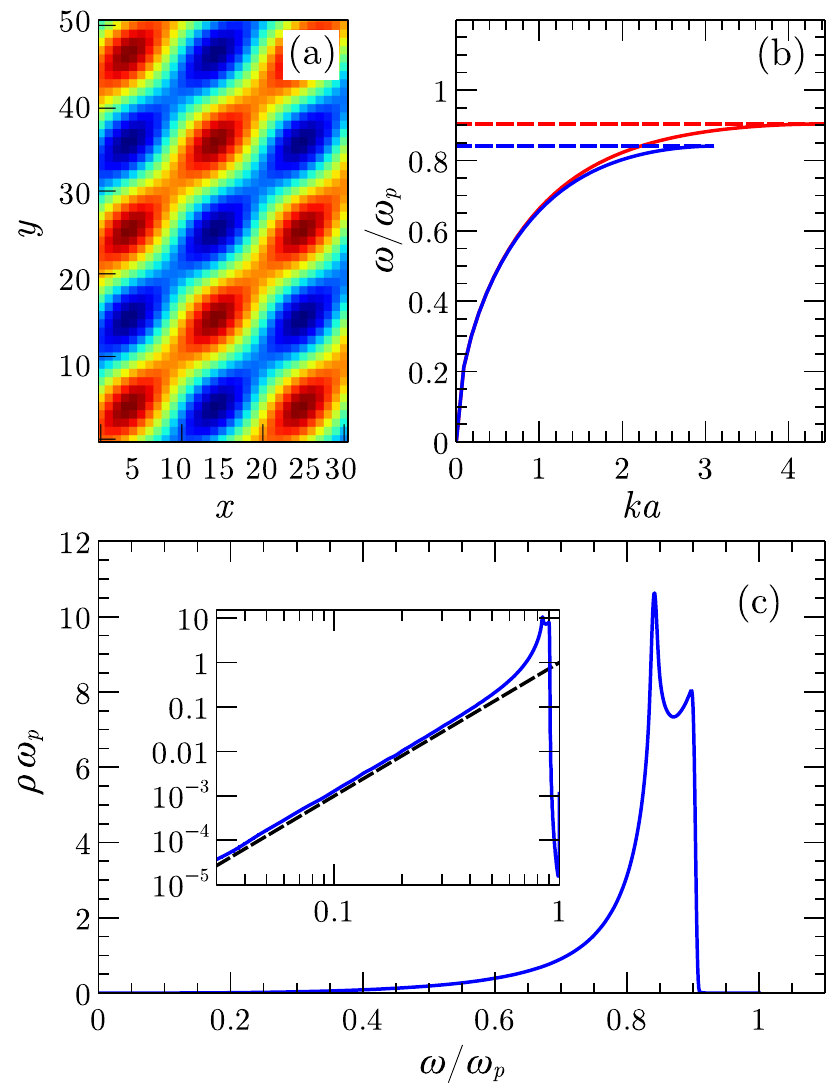}
    \caption{Two-dimensional plasmons in the reduced network model of thin metallic film with filling $p=1.0$. (a) In-plane potential distribution of the eigenmode with resonant frequency $\omega = 0.45 \omega_p$. (b) Dispersion plot for $k$ in the directions $[10]$ (blue line) and $[11]$ (red line) given by Eq.(\ref{eq:2d_plasmon}). (c) The DOS of the reduced network obtained numerically. Inset in the panel (c) shows the DOS in the log-log scale, with the dashed line representing asymptotics from Eq.(\ref{eq:DOS_asymptotics}). Dashed lines in the panel (b) mark frequencies, at which van Hove singularities are present in the DOS. Numerical evaluations are performed with KPM for a network of $10^3 \times 10^3$ sites within $256$ polynomials.}
    \label{fig:2d_plasmon}
\end{figure}

We will first consider resonances in the reduced network with filling fraction of ``metallic'' bonds $p=1.0$. This model corresponds to a very thin metallic film, whose thickness is much smaller than the typical wavelength of the oscillations in the medium. The latter quantity, in its turn, is defined by the network lattice constant $a$. Resonances in such a system have a form of periodic oscillations of charge density in the plane of the resonant layer. An example of eigenmode is presented in Fig.~\ref{fig:2d_plasmon}(a). These resonances form a broad spectrum, ranging from $\omega = 0$ to some limiting value, which we will derive further.

We begin with the analytical derivation of the dispersion relation in order to shed some light on the structure of these resonances. This can be done via substitution of the ansatz $\varphi(x,y,z,t) = \varphi_0{\rm exp}(i\omega t - ik_xx - ik_yy - \varkappa|z|)$ into Kirchhoff's rule $\Sigma_{i,j} I_{ij} =0$ for currents $I_{ij}$ flowing into sites inside and outside the layer. Here $k_x$ and $k_y$ are the components of an in-plane wave vector and $\varkappa$ is an attenuation constant in the $z$ direction. Taking into account Ohm's law and doing necessary calculations, one obtains the dispersion law
\begin{equation}
    \displaystyle \omega^4 = \omega_p^4\frac{{\rm sin}^2(k_xa/2) + {\rm sin}^2(k_ya/2)}{1 + {\rm sin}^2(k_xa/2) + {\rm sin}^2(k_ya/2)}.
    \label{eq:2d_plasmon}
\end{equation}

In the low-frequency domain $k^{-1} \gg a$ this relation resembles a dispersion of the {\it two-dimensional plasmon} in the local response limit $\omega = \omega_p \sqrt{ka/2}$. Indeed, a discreteness of the system does not affect oscillations at the greater wavelengths but plays a crucial role for the resonances with $k^{-1} \propto a$. It is seen from Eq.(\ref{eq:2d_plasmon}) that the group velocity $v_g=\partial \omega/\partial k$ vanishes at the edges of the Brillouin zone $k_{[10]} = \pi/a$ and $k_{[11]}=\sqrt{2}\pi/a$. This results in van Hove singularities, clearly seen in the DOS [Fig.~\ref{fig:2d_plasmon}(c)] at frequencies $\omega = \omega_p(1/2)^{1/4} \approx 0.84 \omega_p$ and $\omega = \omega_p(2/3)^{1/4} \approx 0.9 \omega_p$. The latter frequency defines the high-frequency edge for a resonant band in such lattice systems.

Next, we consider the spectral density of states (DOS)
\begin{equation}
    \displaystyle \rho(\omega) = \frac{1}{N}\sum_j \delta(\omega - \omega_j),
\end{equation}
where $N$ is the number of sites in the reduced two-dimensional network. Numerically calculated DOS is shown in Fig.~\ref{fig:2d_plasmon}(c). The density of states for two-dimensional plasmons with the dispersion $\omega \propto \sqrt{k}$ can be obtained as follows:
\begin{equation}
    \displaystyle \rho(\omega) = \Big(\frac{a}{2\pi}\Big)^{2}2\pi k(\omega)\frac{dk(\omega)}{d\omega} = \frac{4}{\pi}\frac{\omega^3}{\omega_p^4}.
    \label{eq:DOS_asymptotics}
\end{equation}
Comparison of the obtained expression with numerical results reveals a good agreement in the low-frequency region, see Fig.~\ref{fig:2d_plasmon}(c). Thus, in the framework of the new model resonances in a metallic film represent a discrete version of two-dimensional plasmons. At the same time, in the two-dimensional square lattice network with the metal filling $p=1$ all resonances have the same frequency $\omega = \omega_p$ and form the so-called Drude peak. Indeed, in this case, all elements of the matrix $D_{ij}=0$. Hence, all nontrivial eigenvectors of the problem (\ref{eq:GEP}) correspond to the eigenvalue $\lambda=1$. Thus, these resonances represent a quasistatic version of bulk plasmons.

Next, we will  study an influence of a geometrical disorder on plasmon resonances in the reduced network model of thin film composites. To do so, we consider networks with bonds in the plane of the resonant layer being randomly chosen to be ``metallic'' with the probability $p$ or ``dielectric'' with the probability $1-p$. This minimal model is commonly used in papers cited above and corresponds to a limiting case of a discretization lattice with the lattice constant $a$ comparable to the characteristic size of metallic and dielectric inclusions. However, correlations in the local arrangements of bonds have also been considered in Refs.~\cite{1989_Zeng_1, 2001_Zekri}.

\section{Lifshitz tails and a crossover behavior of two-dimensional plasmons}
\label{sec:crossover}

In the present section, we consider the DOS in the reduced model with a metallic bonds filling $p<1$. In this case, $1-p$ metallic bonds are randomly replaced by dielectric ones. If $1-p\ll1$ there are small dielectric holes in a conductive metallic layer. If $p\ll1$ there is a number of small metallic clusters. The percolation threshold does not depend on  a presence of long-range capacitive bonds. Therefore, the percolation threshold is $p_c=0.5$, which is a well-known value for the square lattice.

\begin{figure}[t]
    \includegraphics[width=3.346in]{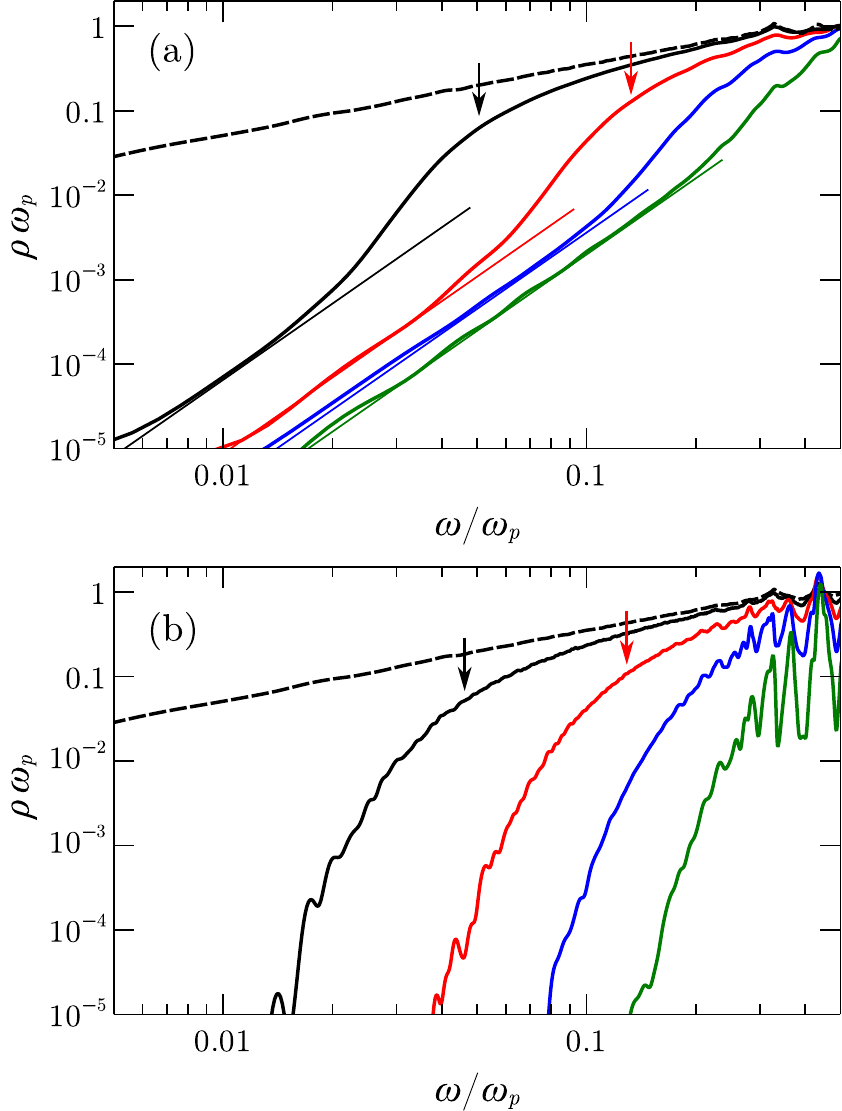}
    \caption{The DOS $\rho(\omega)$ in the reduced network model of a thin film composite for different values of the filling factor $p$: (a) $p=p_c=0.5$ (dashed line) and $p=0.6, 0.7, 0.8, 0.9$ (solid lines from left to right); (b) $p=p_c=0.5$ (dashed line) and $p=0.4, 0.3, 0.2, 0.1$ (solid lines from left to right). The DOS is calculated for a $2000\times2000$ network by the Kernel polynomial method and averaged over $10^3$ realizations. Thin straight lines in the panel (a) show the DOS of two-dimensional plasmons~(\ref{eq:2d_plasmon_DOS_disordered}) for the corresponding values of $p$. Arrows indicate the position of the crossover (a) and the gap width (b) for $|p-p_c|=0.1,0.2$.}
    \label{fig:DOS_Crossover}
\end{figure}

Figure \ref{fig:DOS_Crossover}(a) shows the DOS above the percolation threshold $p\ge p_c$. For $p > p_c$, the low-frequency DOS is similar to the DOS of two-dimensional plasmons: $\rho(\omega) \propto \omega^3$. However, there is a crossover between the low-frequency DOS and the high-frequency one. The crossover frequency $\omega_*$ decreases if $p$ tends to $p_c$. In the case $p=p_c$, there is not any frequency region with $\rho(\omega) \propto \omega^3$. Thus, the crossover is absent, and $\omega_*=0$ for $p=p_c$. In the case $p-p_c \ll 1$, the high-frequency DOS (above the crossover) is close to the DOS for $p=p_c$ [see the case $p=0.6$ in Fig.~\ref{fig:DOS_Crossover}(a)].

Let us consider the low-frequency DOS in detail. For $p=1$, it has the form of Eq.~(\ref{eq:DOS_asymptotics}). However, in the case $p_c < p < 1$, we should take into account that the surface inductance $L_{\rm surf}$ depends on the metallic bonds filling factor $p$, and the effective plasma frequency is proportional to $\sqrt{L_{\rm surf}}$. The surface inductance $L_{\rm surf}$ is proportional to the inductance of each metallic bond $L_m$:
\begin{equation}
    L_{\rm surf}(p) = s(p)L_m.
\end{equation}
Finding an estimate of $s(p)$ is a well-known problem in the percolation theory since $s(p)$ can be treated as a conductivity of the square lattice of unit resistors, which are cut randomly. As a result, the low-frequency DOS has a form
\begin{equation}
    \rho(\omega) = \frac{4}{\pi s^2(p)}\frac{\omega^3}{\omega_p^4}.
    \label{eq:2d_plasmon_DOS_disordered}
\end{equation}
In the vicinity of the percolation threshold $s(p)$ obeys the following scaling relation
\begin{align}
    s(p) &\propto (p-p_c)^t, \quad p>p_c, \\
    s(p) &= 0, \quad p<p_c
\end{align}
with the critical exponent $t=1.3$ \cite{1982_Derrida}. Figure \ref{fig:DOS_Crossover}(a) shows that the low-frequency DOS coincides with Eq.~(\ref{eq:2d_plasmon_DOS_disordered}) where $s(p)$ was obtained via an independent numerical procedure.

The crossover between the low-frequency and the high-frequency DOS can be explained in the following manner. There is a number of dielectric inclusions in the metallic layer (for $p > p_c$). Therefore, plane-wave two-dimensional plasmons can be observed only for wavelengths much greater than the characteristic size $\xi$ of such inclusions. For smaller wavelengths, which are comparable to the size of dielectric inclusions, the structure of resonances strongly depends on a geometry of such inclusions.

Figure \ref{fig:DOS_Crossover}(a) shows that $\rho(\omega)$ near the crossover exceeds the low-frequency DOS given by Eq.~(\ref{eq:2d_plasmon_DOS_disordered}). Therefore, we can qualitatively determine the crossover frequency $\omega_*$, as a position of the maximum in the reduced DOS $\rho(\omega)/\omega^3$. The crossover frequency $\omega_*$ is indicated by arrows in Fig.~\ref{fig:DOS_Crossover}(a) for small values of $|p-p_c|$. Such an excess of the DOS over the low-frequency trend is also known in the vibrational density of states in glasses (the so-called boson peak). Our results emphasize the universality of some properties of disordered systems.

Let us point out that a correct description of resonances requires a study of systems whose size is much larger than the correlation length $\xi$ which diverges at the percolation threshold $p=p_c=0.5$. For instance, at filling $p=0.7$ the crossover is clearly observed in a system with $N\gtrsim10^6$ sites. In this case, $M$ and $D'$ are matrices of the size $[10^6 \times 10^6]$, with $D'$ being a dense matrix. As a result, a direct numerical diagonalization of the generalized eigenvalue problem (\ref{eq:GEP}) is impossible. To this end, we apply the Kernel polynomial method (KPM) \cite{2006_Weisse, 2016_Beltukov}, which allows studying the DOS in such big systems with the help of a modern personal computer.

At fillings $p<p_c$ the $dc$ conductivity is absent. Thus, the system becomes a dielectric one with metallic clusters rather than a conductive one with dielectric inclusions. As a result, no two-dimensional plasmons are present even at low frequencies. Instead, a low-frequency spectral gap with an exponentially small amount of resonances is observed, see Fig.~\ref{fig:DOS_Crossover}(b). The width of this gap increases with a decreasing in $p$. Such a behavior of resonances in disordered systems is known as {\it Lifshitz tails} \cite{1964_Lifshitz}. The low-frequency resonances can be observed only in big clusters of a specific geometry. As shown in Ref.~\cite{1998_Jonckheere}, long chains of metallic bonds in a dielectric environment, the so-called hairpin configurations and worm like configurations act as such clusters in random impedance networks. The probability of such configurations to be observed is exponentially small in the system with a small amount of metallic bonds. As a result, a number of corresponding resonant frequencies is exponentially small as well. A detailed analytical and numerical study of Lifshitz tails in random impedance networks having a geometry of the square lattice is performed in Ref.~\cite{1998_Jonckheere}.

The gap width $\omega_g$ can be obtained using a similar procedure to the one we use for $p>p_c$. We can qualitatively define the gap width as the position of the maximum in the reduced DOS $\rho(\omega)/\omega^3$. The gap width is indicated by arrows in Fig.~\ref{fig:DOS_Crossover}(b) for small values of $|p-p_c|$. One can see that the gap width for $p<p_c$ is approximately the same as the crossover frequency for $p>p_c$ for the same value of $|p-p_c|$. We can estimate a critical behavior of the gap width and the crossover frequency as $\omega_*,\omega_g \propto |p-p_c|^\nu$ with $\nu\approx 1.4$. However, the precise value of $\nu$ is subject to further study.

A rich structure of resonant peaks is observed at low fillings, see  Fig.~\ref{fig:DOS_Crossover}(b) for $p = 0.1$. These peaks correspond to the resonances of small typical clusters, the so-called {\it lattice animals}, which are studied in detail in Refs.~\cite{1996_Clerc, 1998_Jonckheere}.

Let us point out that in previously studied square lattice networks the DOS $\rho(\omega, p) = \rho(\omega, 1-p)$ due to self-duality of this lattice \cite{1998_Jonckheere, 1971_Dykhne}. As a result, in such networks, peaks associated with lattice animals as well as Lifshitz tails are present both at fillings $p<p_c$ and $p>p_c$. Thus, the reduced model demonstrates a qualitative difference with respect to previously studied two-dimensional impedance networks, namely the presence of low-frequency two-dimensional plasmon like waves.

\section{Conclusion}
\label{sec:conclusion}

In the present paper, we introduced a random impedance network model which describes plasmon resonances in disordered thin film materials. This model takes into account the two-dimensional geometry of the resonant layer carrying plasmonic inclusions as well as the three-dimensional electrodynamics of the interaction between them. The corresponding network can be reduced to the two-dimensional one with capacitive bonds connecting distant sites. The capacitance of these bonds $C_{ij}$ decreases with the distance between sites $i$ and $j$ as $C_{ij} \propto 1/r_{ij}^3$.

In order to analyze the resonant spectrum of the considered system, we applied the Kernel polynomial method. It allows carrying out numerical studies of random networks with several millions of sites in the resonant layer. We have considered the simplest model, in which an arbitrary bond between nearest neighbors in the resonant layer can be metallic with probability $p$ or dielectric with probability $1-p$.

The density of low-frequency resonances in the considered model at fillings $p>p_c$ is analogous to the density of two-dimensional plasmons. However, plane-wave two-dimensional plasmons can be observed only at wavelengths much greater than the characteristic size of dielectric inclusions in the metallic layer. For higher frequencies and smaller wavelengths, which are comparable to the size of dielectric inclusions, a structure of resonances is controlled by the geometry of such inclusions. The DOS near the crossover supersedes the density of two-dimensional plasmons. This can be observed as a peak in the reduced DOS $\rho(\omega)/\omega^3$. This peak is analogous to the boson peak, a well-known phenomenon in physics of glasses. A more detailed study of the observed crossover is a subject for future work.

At the same time, Lifshitz tails are present at low frequencies for fillings $p<p_c$, as in previously considered models with the two-dimensional electrodynamics. Characteristic frequencies of the crossover and of Lifshitz tail are approximately equal for the corresponding values of $|p-p_c|$.

We would like to stress a substantial difference between the introduced model and those considered in the previous contributions we are aware of. It is the presence of long-range capacitive connections in the network, whose effect is mathematically equivalent to the presence of the three-dimensional cubic lattice of capacitors that surrounds the two-dimensional resonant network. This lattice changes the electrostatic interaction in the network in such a way that the charge potential is $\varphi \propto 1/r^2$ instead of $\varphi \propto 1/r$ observed in the two-dimensional square lattice network. The latter is characteristic of the two-dimensional electrostatics, whereas the first one is more typical for three-dimensional problems. Such a difference in potentials leads to a significant modification of key features of resonances in the model network, namely to the different dispersion law of plasmons and the presence of new phenomena in disordered networks, including the crossover between two-dimensional plasmons and resonances of small clusters, as well as the absence of symmetries associated with the self-duality.

Our approach provides a simple model capable of a qualitative description of plasmon resonances in disordered thin film nanocomposites near the percolation threshold, as it allows studying phenomena assisted by long-range correlations in the composite geometry. A role of such correlations in light emission by embedded fluorescent sources has been analyzed in recent experiments with semicontinuous gold films and plasmonic lithographic networks \cite{2015_Gaio}. Measurements of optical absorption in disordered gold films \cite{2014_Hedayati} demonstrate particular features of our model, namely the presence of low-frequency absorption tails at high fillings that are absent in continuous films and in films at low fillings, the presence of the maxima in the absorption for particular filling values as well as the presence of the optimal filling which maximizes the absorption, and the absence of the self-duality.

The model can be easily generalized to study materials with arbitrary constituents. This can be done via a substitution of the desired dielectric functions into Eq.(\ref{eq:lambda}) without any additional numerical simulations \cite{1998_Jonckheere, 2016_OBP}. For instance, resonances in binary networks composed of two dissimilar metallic bonds have been considered \cite{1997_Baskin, 2016_OBP}, as well as properties of three-component networks \cite{2003_Gu, 2004_Gu_1} and networks with continuously distributed values of bonds \cite{1992_Bergman, 2015_OBP}.

\section*{Acknowledgments}
We are very grateful to V.I. Kozub, D.A. Parshin, A.L. Efros and C.R. Simovski for fruitful discussions. This work was financially supported by the Russian Foundation for Basic Research (Project no. 16-32-00359), the ``Dynasty'' Foundation, and the Government of the Russian Federation (Grant 08-08).

\begin{figure}[b]
    \includegraphics[width=3.346in]{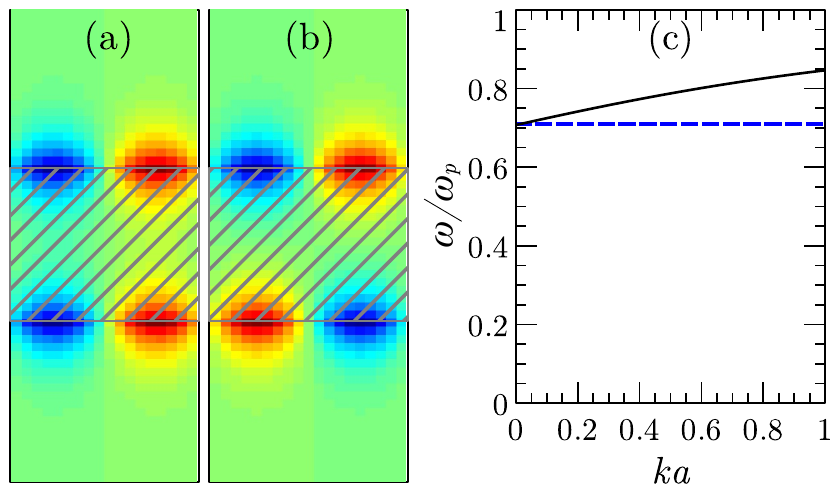}
    \caption{Electric potential distribution in the symmetric (a) and antisymmetric (b) plasmon modes of a metallic film. The film and the surrounding dielectric are modelled by the network of $60 \times 60$ sites that represents the cross section of the system in the plane perpendicular to the film. The metallic region consists of $20$ resonant layers and is shown with grey hatching. (c) The dispersion relation for surface plasmons at the metal-dielectric interface. Solid line represents the analytic result for the network model (\ref{eq:SPP_dispersion}), dashed line shows the quasistatic prediction $\omega(k) = \omega_p/\sqrt{2}$.}
    \label{fig:SPP}
\end{figure}

\section*{Appendix A. Limitations and possible extensions of the model}
In the present Appendix, we discuss the approximations of the considered network model in more details. First, the model is based on the quasistatic approach. However, for disordered systems composed of deeply subwavelength inclusions, this does not lead to a sufficient difference in the general picture with respect to the consideration that takes into account retardation effects. In particular, a corresponding generalization of the two-dimensional network model is considered in Refs.~\cite{2000_Sarychev, 2000_Shubin}. It is shown that the square lattice random impedance network is replaced by two square lattice networks. Voltages at bonds of one network describe electric fields in the composite, and another network describes magnetic fields in a similar manner. In this case, strong fluctuations of the electric field remain \cite{2000_Sarychev}. Magnetic fields also demonstrate the presence of hot spots and giant fluctuations \cite{2000_Sarychev}.

Second, we considered an example of very thin systems which can be described with a single network layer composed of resonant bonds as well as dielectric ones. This layer is surrounded by the three-dimensional simple cubic lattice of dielectric bonds that corresponds to the surrounding vacuum and carries displacement currents. As a result, we obtain two-dimensional plasmons with the dispersion $\omega \propto \sqrt{k}$ in the system with metal filling $p=1$. However, such oscillations are rather observed in graphene and other two-dimensional materials \cite{2014_Yoon}, whereas for metal films the presence of {\it surface plasmon-polaritons} (SPP's) is a much more typical situation \cite{Maier-book}. In thin films, there are two types of surface plasmon modes, symmetric and antisymmetric, see Figs.~\ref{fig:SPP}(a) and \ref{fig:SPP}(b). Also, SPP's have a different dispersion relation, which in the quasistatic approach reduces to the constant frequency of the surface plasmon resonance (SPR) $\omega(k) = \omega_p/\sqrt{1 + \varepsilon_d}$ for both the symmetric and the antisymmetric mode, as well as for the surface plasmon at the metal-dielectric interface [Fig.~\ref{fig:SPP}(c)]. The dispersion of resonances in the random impedance network that corresponds to the metal-dielectric interface can be easily obtained in a similar way as Eq.~(\ref{eq:2d_plasmon}) and reads as
\begin{equation}
    \displaystyle \omega^2 = \frac{\omega_p^2}{1 + \varepsilon_{d}} ( 1 + \frac{{\rm sin}(ka/2)}{\sqrt{1 + {\rm sin}^{2}(ka/2)}}).
    \label{eq:SPP_dispersion}
\end{equation}
However, this dispersion describes only the influence of the discreteness of the lattice, which is seen clearly at wavelengths comparable to the lattice constant $ka \approx 1$ and is almost absent at large wavelengths $ka \ll 1$. The construction of a reduced model seems to be not so straightforward in this case. Thus, a study of plasmon resonances in such ``thick'' disordered films remains an open task.

We would also mention some modern papers that study plasmon resonances in disordered systems via full-wave numerical simulations \cite{2010_Chettiar, 2017_Frydendahl, 2017_Toranzos}. Such an approach requires significant computational resources with respect to impedance network models and thus does not allow studying large-scale properties of resonances in systems near the percolation threshold. Yet, numerical modeling seems to be the most precise method to study local effects, e.g., hot spots in small arrangements of nanoparticles.

\end{document}